\documentstyle[12pt,aaspp]{article}
\def\kms{{\rm km/s}}
\begin{document}

\def\thefootnote{\fnsymbol{footnote}}

\title{THE OPTICAL GRAVITATIONAL LENSING EXPERIMENT. \\
OGLE \#7:  BINARY MICROLENS OR A NEW UNUSUAL
VARIABLE?\footnote{Based on observations obtained at the Las
Campanas Observatory of the Carnegie Institution of Washington}
}

\def\thefootnote{\arabic{footnote}}

\author{A. Udalski\altaffilmark{1}, M. Szyma\'nski\altaffilmark{1},}
\affil{\tt e-mail I: (udalski,msz)@sirius.astrouw.edu.pl}
\author{S. Mao\altaffilmark{2}, R. Di~Stefano\altaffilmark{2},}
\affil{\tt e-mail I: (smao,rd)@cfata3.harvard.edu}
\author{J. Ka\l u\.zny\altaffilmark{1}, M. Kubiak\altaffilmark{1},
M. Mateo\altaffilmark{3}, W. Krzemi\'nski\altaffilmark{4}}
\affil{\tt e-mail I: (jka,mk)@sirius.astrouw.edu.pl,
mateo@astro.lsa.umich.edu, wojtek@roses.ctio.noao.edu}

\altaffiltext{1}{Warsaw University Observatory, Al. Ujazdowskie 4,
00--478 Warszawa, Poland}
\altaffiltext{2}{MS 51, Center for Astrophysics, 60 Garden Street,
Cambridge, MA 02138}
\altaffiltext{3}{Department of Astronomy, University of Michigan, 821 Dennison
Bldg., Ann Arbor, MI 48109--1090}
\altaffiltext{4}{Carnegie Observatories, Las Campanas Observatory, Casilla
601, La Serena, Chile}

\begin{abstract}

We present the light curve of an unusual variable object, OGLE~\#7,
detected during the OGLE search for microlensing events. After one
season of being in a low, normal state, the star brightened by more than
2~mag with a characteristic double-maximum  shape, and  returned to
normal brightness after 60 days. We consider  possible explanations of
the photometric behavior of OGLE~\#7. The binary microlens model seems
to be the most likely explanation -- it reproduces well the observed
light curve and explains the observed colors of OGLE~\#7. The
characteristic time scale of the OGLE~\#7 event, $t_E$, is equal to 80
days, the longest observed to date. The binary microlens model predicts
that the spectrum of the star should be composite, with roughly 50\% of
its light in the $I$-band coming from a non-lensed source.

\end{abstract}


\clearpage

\section{INTRODUCTION}

The Optical Gravitational Lensing Experiment (OGLE) is a long term
observing  project with the main goal of searching for dark  matter in
our Galaxy using microlensing (Paczy\'nski 1986). After two years of
continuous monitoring of approximately two million non-variable stars in
the direction of the Galactic bulge, ten microlensing events have been
detected (Udalski et al. 1993b, Udalski et al. 1994a,  Udalski et al.
1994b). The resulting optical depth of microlensing  $(3.3 \pm 1.2)
\times 10^6$ is higher than previous theoretical estimates and indicates
that the bulge of our Galaxy is in fact a bar, with its long axis
inclined $ \sim 15 $ degrees to the line of sight (Paczy\'nski et al.
1994b).

A microlensing event caused by a single, point-like lensing object has
an achromatic light curve  with a well defined, characteristic shape
which is symmetric around the maximum of brightness (Paczy\'nski 1986,
1991). The lensing phenomenon should  not repeat as the probability of
the same star being lensed twice is negligible. All the OGLE
microlensing events except OGLE~\#7 displayed such photometric behaviors
(for OGLE \#6, a better fit to the data can be obtained with a binary
microlens model, however, the case is not strong, Mao \& Di~Stefano
1994). All the reported MACHO (Alcock et al. 1994) and EROS (Aubourg  et
al. 1993) events can also be well fit by single lenses.

It is well known that majority of stars are found in binary or multiple
systems (Abt 1983).  Therefore it is natural to expect some microlensing
events to be caused by binary systems. The light curve of such events
might be considerably different from that of single microlens events and
would depend on the geometry of lensing (Mao \& Paczy\'nski 1991, Mao \&
Di~Stefano 1994). Light curve may also exhibit  color variations, although
these may be difficult to detect because of small amplitude.

One of the OGLE microlensing candidates found during the OGLE search,
OGLE \#7, exhibits very unusual light variations which do not resemble
variations of any known variable stars. On the other hand, the light
variations are strikingly similar to the theoretical light curves
possible for binary microlenses (Mao \& Paczy\'nski 1991). In this {\it
Letter} we present a detailed analysis of the variability of OGLE \#7
candidate, emphasizing binary microlensing as a possible explanation of
observed light variations. Section 2 describes details of the
observations, Section 3, the OGLE \#7 light curve and Section 4,
possible explanations of its photometric behavior.  We conclude the
paper with a discussion in Section 5.

\clearpage

\section{OBSERVATIONS}

The OGLE microlensing search is conducted using the 1-m Swope telescope
at Las Campanas Observatory which is operated by Carnegie Institution
of Washington. A $2048 \times 2048$ pixels Loral CCD chip, giving the
scale of 0.44 arcsec/pixel at 1-m telescope focus, is used as a
detector. Collected frames feed the data pipeline which, almost in real
time, derives the photometry of objects. The main search is conducted in
the $I$-band with some frames occasionally  taken  in the $V$-band.
Details of the reduction techniques are described in Udalski et al.
(1992).

OGLE \#7 candidate: the Baade's Window field  BW8 (Udalski et al. 1992)
star $I$~198503 -- has been detected during the regular microlensing
search as described in Udalski et al. (1993b) and Udalski et al.
(1994a). It was also detected in a fully algorithmic search described by
Udalski et al. (1994b), although that detection could have been
accidental since the OGLE \#7 light curve is different from the regular
microlensing light curve used in the algorithm.

To minimize the photometric errors  resulting from instrumental errors
and  from the general technique of reductions used in the OGLE search,
the final differential photometry of OGLE \#7  with respect to two
bright, nearby stars was derived as in Udalski et al. (1994a).
Differential photometry was then tied to the standard system. Table 1
contains the position and colors of OGLE~\#7 and the comparison stars.
A $45 \times  45$ arcsecs region centered  on the star is shown in
Figure~1.

\section{LIGHT CURVE OF OGLE \#7}

Figure 2 shows the light and $V-I$ color curves of OGLE \#7 from the
1992 and 1993 observing seasons. The error bars correspond to the errors
given by the photometry code DoPhot (Schechter, Mateo, and Saha 1993)
rescaled by 1.4  to approximate the real errors of observation (Udalski
et al. 1994b).

During the 1992 observing season the brightness of OGLE \#7 remained at
a constant level with $I \approx 17.5$ mag and the star therefore was
included in the non-variable object database for the microlensing
search. Dramatic light changes occurred in the next (1993) observing
season. At the beginning of the 1993 observing season the star was still
close to its previous low light level: $I \approx 17.3$. Then it
brightened gradually by about 0.2 mag. Around day $\hbox{JD} -
2448000=1140$  it rose  rapidly by about 2 mag reaching $I \approx
15.1$. Unfortunately, due to telescope scheduling constraints and
unusually bad weather, that part of the light curve is sparsely covered
by observations. Moreover the observation taken at the maximum light was
obtained in poor seeing conditions. Nevertheless, due to the large
brightness of OGLE~\#7 at that time, the derived photometry is reliable.
After reaching the maximum, the brightness of OGLE~\#7 faded and reached
a plateau about 1 mag brighter than the minimum level. Then it began to
rise again, followed by an abrupt  drop to the level 0.3 mag over the
minimum level. After that, OGLE~\#7  gradually faded to minimum
brightness.  The overall light variations  have a double-maximum
structure with `U'-shape changes between them. Very smooth light changes
during the event  indicate lack of short time scale variations exceeding
a few hundreds of a magnitude.

It is worth noting that the OGLE~\#7 candidate is being continuously
monitored during the OGLE 1994 observing season. Up to the time of
writing this paper, the brightness of the star has remained at the low
level with $I \approx 17.5$ mag over the entire period
$\hbox{JD}-2448000$: 1439 -- 1560.

OGLE~\#7 was monitored mainly in the $I$ filter. From six $V$-band
observations the $V-I$ color of the star is $V-I=1.94 \pm 0.07$. There
is no information about color during the event. However, the observation
taken after the event when the star was still about 0.3 mag above the
minimum suggests  no significant color variation.

\section{WHAT IS OGLE \#7?}

The photometric light curve of the OGLE~\#7 indicates that we have
discovered a very mysterious object with unusual properties. In this
Section we explore its true nature.

\subsection{An Eruptive Variable?}

The rapid increase of brightness might suggest some kind of eruptive
variable as a possible explanation of the OGLE~\#7 behavior. The
amplitude of   brightening excludes any type of cataclysmic variables
other than dwarf novae or  low-mass X-ray binaries (LMBX). However, it
seems unlikely that OGLE~\#7 is a dwarf nova. First, the light curve
does not resemble that of dwarf novae, and the duration of the event is
much longer than typical dwarf novae outbursts or superoutbursts.
Second,  the $V-I$ color of OGLE~\#7 (Table~1), $V-I=1.9$, is too red
for a typical dwarf nova system even after taking into account
extinction toward the bulge (upper limit: E$_{V-I}=0.55$ mag,
Paczy\'nski et al. 1994a).

The remaining option among cataclysmic variables -- a low mass X-ray
binary -- also does not seem likely due to essentially the same
arguments. The light curve does not resemble that of a low mass X-ray
binary in a high state. It is very smooth, which indicates no light
changes larger than a few hundreds of magnitude whereas LMXBs usually
exhibit significant variations caused by reflection effects and/or by
flickering. X-ray novae -- a subgroup of X-ray binaries -- can be
excluded because  the amplitude of OGLE~\#7 is too small. The $V-I$
color seems to be  also too red for a typical LMBX star. Moreover, we
are not aware of any detection of an X-ray source when OGLE~\#7 was in a
high state, nor of any known X-ray source close to the position of
OGLE~\#7.

Another possibility could be a flaring late spectral type star. However,
typical flares in such stars last minutes or hours, and we do not know
any object with flares lasting for several days. Also the double-maximum
light curve does not resemble that of flaring stars. With such long
optical coverage of OGLE~\#7 we would expect at least some traces of
another flaring activity but we have not registered any other light
increase except the event described above.

The optical spectrum of OGLE~\#7 could shed some light onto the nature
of the object. As cataclysmic variables and flaring stars often reveal
emission lines, their presence would be an argument in favor of an
eruptive nature for the object. Also, if the observed brightening is due
to some eruptive processes, it should repeat in the future.

\subsection{A Binary Lens?}

Lensing by binaries is naturally expected in $\sim 10\%$ of the OGLE
events (Mao \& Paczy\'nski 1991). It is thus statistically plausible to
have one binary lensing event in the ten OGLE events sample (Udalski et
al. 1994b). Furthermore, the characteristic `U'-shape light variations
between maximas in  OGLE~\#7's high state can be found in the
theoretical light curves of binary lenses (e.g., Mao \& Paczy\'nski
1991). The `U' shape is a result of the source crossing the caustics.
Another important argument in favor of the binary lens interpretation is
provided by the colors of OGLE~\#7, which indicate that the star is a
typical Galactic bulge main sequence turn-off point star (Udalski et al.
1994b) and thus can be a source for microlensing. All the above
considerations lead us to check whether a binary microlens model can
reproduce the observed light curve.

The fitting of a binary light curve has been studied in some detail by
Mao \& Di~Stefano (1994). The fitting involves the search for a global
minimum of a $\chi^2$ measure in multiple dimensional parameter space.
For the minimal binary model, there are seven parameters: i.e.,  the
total mass of the binary, $M$; the mass ratio of the individual masses,
$q=m_2/m_1$; their separation, $a$, as projected onto the lens plane
(defined in terms of the center of mass of the binary); the closest
approach to the center of mass, $b$; the time of closest approach,
$t_b$; the angle between the axis of the binary and the trajectory,
$\theta$; and, the magnitude of the star at the minimum light, $I_0$.
For the Galactic bulge observations, we need one additional parameter,
$f$, the light contributed by the lensed star at  minimum of light. This
parameter is necessary to take into account the light contribution of
the lens(es) and/or a nearby star located close to the source with
separation smaller than the seeing disk ($\la 1$ arcsec). For the
technical details of the fitting procedure, the readers should refer to
Mao \& Di~Stefano (1994).

In Fig.\ 2 we plot our best fit model. The model has a $\chi^2$ (as
defined in Mao \& Di~Stefano 1994) of  139 for 83 data points (75 degree
of freedom). The $\chi^2$ for the varying part (between days
$\hbox{JD}-2448000$: 1085 -- 1240) of the light curve is roughly 57 for
51 data points. The best fit parameters are, respectively,
$$q=1.02;~a=1.14 R_E;~b=0.050 R_E; ~ \theta=138^\circ.3; $$
\begin{equation} \label{model}
{}~ t_E=80~\hbox{day}; ~ t_b=1172.5~\hbox{day}; ~ f=56\%; ~ I_0=18.1,
\end{equation}
where $R_E$ is the Einstein radius projected onto the source plane for
the total mass $M$. The lensing geometry is  shown in Fig.\ 3. We have
assumed that the lensed star is a point object, and ignored the circular
component of the motions of the lenses and the Earth on the scale of a
few AU. The validity of these assumptions will be checked below.

The best fit indicates that about half of the OGLE~\#7 light in the
normal state must come from an additional source. There are two obvious
possibilities for the additional light: a very close optical companion
of the lensed star which cannot be separated from the ground, or, the
binary lensing system itself.

  From the composite colors observed in the low, normal state we can put
some limits on the components of such a blend. The combined magnitudes
of the low state are $V=19.4$ and $I=17.5$. Thus the additional light
must have $I=18.4$ while the lensed star $I=18.1$. Assuming that the
lensed star should be in the Galactic bulge its $V-I$ color should not
exceed 2.5 mag (see CMD in Udalski et al. 1993a). Thus we can set a
lower limit on the colors of the source of the additional light: $V >
19.8$ and $V-I > 1.4$, and an upper limit on its intrinsic color:
$(V-I)_0 > 0.9$, where we have corrected the extinction for the Baade's
Window field BW8,  $E(V-I)=0.55$ (Paczy\'nski et al. 1994a)

If the additional light comes from the binary lensing system itself,
both components of the system should have similar spectral type, as the
mass ratio is close to one. Thus the color limitation indicates that
both components have spectral type later than K1 and their masses must
be smaller than $0.8 M_{\odot}$, assuming normal, solar metallicity main
sequence stars. On the other hand, two K1 type stars cannot be further
from us  than  4 -- 5 kpc to have the observed $V$ luminosity (assuming
1 -- 1.2 mag extinction). Fainter objects must have been located much
closer. For example, early M type components would be only 1 -- 2 kpc
from us -- a distance that makes lensing very unlikely. Taking both
limits together we conclude that if the lensing binary system is the
only source of additional light then it must consist of two K-type stars
with masses $0.5 - 0.8 M_{\odot}$ located in the Galactic disk at a
distance of 3--5 kpc.

The second possibility for the additional light is a very close ($<0.5$
arcsec) optical companion of the lensed  star. In this case we can  put
only the following constraint on the lensed star: $V=19.4$, $V-I=1.3$ if
the companion is very red and does not contribute to the $V$-band. The
most likely situation is a bulge companion with color $V-I=1.6$ which
gives $V=20.3$ and $V-I=2.2$. In either case the lensed star is a
typical Galactic bulge star (Udalski et al. 1993a).

We now check whether the assumption of the lensed star being a point
object is valid. The lensed star $I$-band magnitude is $I=18.1$, and it
must lie on the line ($V=19.4, V-I=1.3$) -- ($V=20.6, V-I=2.5$) on the
CMD.  Following Paczy\'nski et al. (1994a), the extinction  to the BW8
field and color excess are  $A_V=1.4$, $E(V-I)=0.55$, respectively. The
absolute $V$ magnitude and intrinsic color of the source can be
estimated to be in the range  $M_V=3.5, (V-I)_0=0.75$ to $M_V=4.7,
(V-I)_0=1.9$, where we have assumed that the distance to the bulge is 8
kpc. That corresponds to a slightly evolved G-M spectral type main
sequence star. The upper limit for the radius of such a star can be
safely set to be $R_\star=10 R_\odot=7\times 10^{11}~{\rm cm}$. We can
now  compare this with the Einstein radius, which is given by
\begin{equation}  R_E=V_t ~ t_E = 1.4 \times 10^{14}~{\rm cm}~
{V_t \over 200~ \kms},
\end{equation}
where $V_t$ is the total transverse velocity of the lens, observer and
source projected onto the source plane. The source radius is at least
one order of magnitude smaller than the inter-caustic spacing (see
Fig.\ 3). The point source approximation is excellent except for the
points close the fold caustics. For these points, the source size sets a
limit on the maximum magnification, $A_{\rm max} \approx
(R_\star/R_E)^{-1/2}$. For OGLE \#7, this does not affect the fit. This
can also be seen in Fig.\ 2, where the light curve shown is obtained for
a star with radius $R_\star=5 R_\odot$. The resulting light curve fits
the data as well as that of a point source.

As the Einstein radius is related to the mass of the lens, we can
estimate the total mass of the binary lens to be
\begin{equation} \label{M}
M=1.3 M_\odot \left ({ V_t \over 200~\kms} \right)^2 \left( {  x \over
1-x } \right),
\end{equation}
where $x=D_L/D_S$ is the ratio of the distances to the lens and the
lensed star. We have shown that if the additional light comes from the
binary lens, then the lensing object must be located in the disk, about
4 kpc from us ($x=0.5$) where the most likely velocity is about 200
\kms. Eq. (\ref{M}) then gives a total mass of $1.3M_{\odot}$, i.e.,
about $0.7M_{\odot}$ for each component in excellent agreement with the
color constraints.

The circular velocity of the binary can be easily calculated to be $V_b
\la 10~\kms$, an order of magnitude smaller than the expected transverse
velocity. Also, the circular motion of the Earth around the Sun is
small, $v_\oplus=30~\kms$, compared with the transverse velocity. Thus
both can be ignored.

\section{DISCUSSION}

In the previous section we considered possible explanations of the
strange photometric behavior of OGLE~\#7. It seems that the binary
microlens model is the most likely explanation, as the model reproduces
well the light variations and explains the observed colors of the star.
However, we cannot rule out the possibility that we have  discovered a
completely new type of variable star, and that the binary microlens
model is simply a false positive (cf. Mao \& Di~Stefano 1994). However,
such a false positive seems unlikely.

Generally speaking, there may be multiple binary lens fits to a given
light curve, especially one with weak (small amplitude) lensing
signature (Mao \& Di~Stefano 1994). Fortunately, OGLE \#7 is a very
strong event (brightening by more than two  magnitudes) with distinctive
features; correspondingly the models are well constrained.  The best
model presented here is  significantly better than any other fits we
have found. We believe that it is probably the unique solution in the
multiple dimensional parameter space. Our model not only provides a good
fit to the somewhat sparsely sampled light curve, it also makes a very
specific prediction, i.e., the spectrum should be a composite of at
least two sources of comparable brightness. If the extra light in our
model is contributed by an additional source close to the lensed star,
then the additional source is likely to have a different radial velocity
from the lensed star. The velocity difference may not be large,
therefore high dispersion instruments (e.g., Keck telescope) are needed.
On the other hand, if the binary lens contributed  the extra light, then
the circular motion of the two components ($\sim 10\kms$) can be
measured spectroscopically. This will in turn reliably determine the
other parameters (including the lens mass) in our model. There is one
more test which distinguishes the lensing model from other kinds of
variability. The microlensing event should not repeat, as the
probability of the same star being microlensed more than once is
practically zero. Thus if brightening of OGLE~\#7 repeats, then  the
microlens model will have to be rejected. To summarize, all the
observations are fully consistent with the binary microlens model; our
model can be further tested by high dispersion spectroscopic
observations from the ground, by  high resolution imaging with HST and
by future long-term monitoring of the source. If future observations
contradict the prediction of the binary scenario, then more elaborate
models will have to be invented, a possibility we choose not to pursue
here.

The theoretical estimate that 10\% of the events may exhibit binary
features (Mao \& Paczy\'nski 1991) was based on the binary population
study of the solar neighborhood. However, it is conceivable that the
fraction could be higher if the stellar population in the bulge is
different. Binary light curves are usually very bright (OGLE \#7 reaches
15~mag in the $I$ band), therefore these events can be more easily
detected than single lens events. With the implementation of the early
warning system of OGLE (Paczy\'nski 1994), most of these binary events
can be caught in the early stages and therefore be well sampled in the
bright caustic crossing phase. For the microlens model of OGLE \#7, the
caustic crossing lasted for only $5 \times R_\star/(5R_\odot)$ hours.
This indicates that frequent sampling is extremely important. A densely
sampled caustic crossing event will allow us to resolve the structure of
the source more easily than for the case of a single lens (Gould 1994;
Nemiroff \& Wickramasinghe 1994; Witt \& Mao 1994). In addition,
relatively small changes in the lensing parameters will induce large
change in the light curve; therefore, a well sampled binary lens light
curve offers the hope that all of the parameters can be determined
accurately, including the mass of the lenses.

Photometry of OGLE~\#7 and other microlensing events, as well as a
regularly updated OGLE status report can be found over the Internet at
host ``sirius.astrouw.edu.pl'' (148.81.8.1), using the ``anonymous ftp''
service (directory ``ogle'', files ``README'', ``ogle.status'', ``early
warning''). The status report contains the latest news and references to
all the OGLE related papers, and the PostScript files of some
publications. Information on the recent OGLE status is also available
via  ``World Wide Web'' WWW: ``http://www.astrouw.edu.pl/''.

\acknowledgments{It is a great pleasure to thank B. Paczy\'nski for
valuable suggestions and discussions. This project was supported with
the NSF grants AST 9216494 and AST 9216830 to B. Paczy\'nski and Polish
KBN grants No 2-1173-9101 and BST475/A/94 to M. Kubiak.}

\clearpage

\pagestyle{empty}

\begin{small}
\begin{planotable}{lccc}
\tablewidth{35pc}
\tablecaption{OGLE~\#7 and Comparison Stars }
\tablehead{
\colhead{Parameter} & \colhead{OGLE~\#7} & \colhead{Comparison A}  &
\colhead{Comparison B} }
\startdata
RA$_{2000}$  &   18:03:35.74 & 18:03:34.58 & 18:03:37.18 \nl
DEC$_{2000}$ &  --29:42:01.3 &--29:42:08.6 &--29:41:57.6 \nl
$V$          &    19.4       &   17.45     &  17.57      \nl
$V-I$        &     1.9       &    2.11     &   2.12

\end{planotable}
\end{small}
\clearpage
%
%
%
\clearpage
\begin{figure}
\caption{
45 by 45 arcsec $I$-band images centered on the OGLE \#7 star. Two arrows
identify the lensed star. North is up and east is to the left.
}
\end{figure}

\begin{figure} \caption{
The light and $V-I$ color curves of the OGLE \#7. The error bars
correspond to the formal errors returned by the DoPhot photometry
program rescaled by 1.4 to approximate errors of observations (Udalski
et al. 1994b). The solid line shows the light curve of the best binary
microlens model. Two blown-ups (a,b) show details of the light curve
around two peaks in the model light curve. Although the model parameters
(eq. [1]) are found using a point source approximation, the light curve
shown here is obtained by convolving the point source light curve with
that of a star with radius $R_\star=5 R_\odot$ and a ``cosine'' limb
darkening profile in the $I$-band. We adopt a transverse velocity of
$200 ~ \kms$. The dashed line indicates the magnitude of the extra light
provided by the binary lens and/or an unresolved nearby source.
}
\end{figure}

\begin{figure}
\caption{
The geometry of the best fit binary microlens model. The two components
are labeled as two black dots. The caustics (thick solid) and the
critical curves (dashed) are both projected onto the source plane. The
trajectory of the source is indicated as a straight line with an arrow
indicating the direction of the motion. Both axes are in units of the
Einstein radius ($R_E \approx 10$ AU) in the source plane.
}
\end{figure}


\clearpage
\begin{references}
%
\reference Abt, H.A.  1983, ARA\&A, 21, 343.

\reference Alcock, C. et al. 1994, ApJL, submitted

\reference Aubourg, E. et al. 1993, Nature, 365, 623

\reference Gould, A. 1994, \apj, 421, L71

\reference Mao, S., \& Paczy\'nski, B. 1991, \apj,  374, L37

\reference Mao, S., \& Di~Stefano 1994, \apj, submitted

\reference Nemiroff, R. J., \& Wickramasinghe, W. A. D. T. 1994,
	\apj, 424, L21

\reference Paczy\'nski, B. 1986, ApJ, 304, 1

\reference Paczy\'nski, B. 1991, ApJ, 371, L63

\reference Paczy\'nski, B. 1994, IAU Circ. 5997

\reference Paczy\'nski, B., Stanek, K. Z., Udalski, A., Szyma\'nski, M.,
Ka\l u\.zny, J., Kubiak, M., \& Mateo, M. 1994a, AJ, 107, 2060

\reference Paczy\'nski, B., Stanek, K. Z., Udalski, A., Szyma\'nski, M.,
Ka\l u\.zny, J., Kubiak, M., Mateo, M., \& Krzemi\'nski, W. 1994b,
ApJL, submitted

\reference Schechter, P.L., Mateo, M., and Saha A. 1993, P.A.S.P., 105,
1342.

\reference Udalski, A., Szyma\'nski, M.,
Ka\l u\.zny, J., Kubiak, M., \& Mateo, M. 1992, Acta Astron., 42, 253

\reference Udalski, A., Szyma\'nski, M.,
Ka\l u\.zny, J., Kubiak, M., \& Mateo, M. 1993a, Acta Astron., 43, 69

\reference Udalski, A., Szyma\'nski, M., Ka\l u\.zny, J., Kubiak,
M., Krzemi\'nski, W., Mateo, M., Preston, G. W., \& Paczy\'nski, B.
1993b, Acta Astron., 43, 289

\reference Udalski, A., Szyma\'nski, M., Ka\l u\.zny, J., Kubiak, M.,
Mateo, M., \&  Krzemi\'nski, W. 1994a, ApJL, 426, L69

\reference Udalski, A., Szyma\'nski, M., Stanek, K. Z., Ka\l u\.zny, J.,
Kubiak, M., Mateo, M., Krzemi\'nski, W., Paczy\'nski, B., \& Venkat, R.
1994b, Acta Astron., 44, 165.

\reference Witt, H., \& Mao, S. 1994, \apj, in press
%
\end{references}
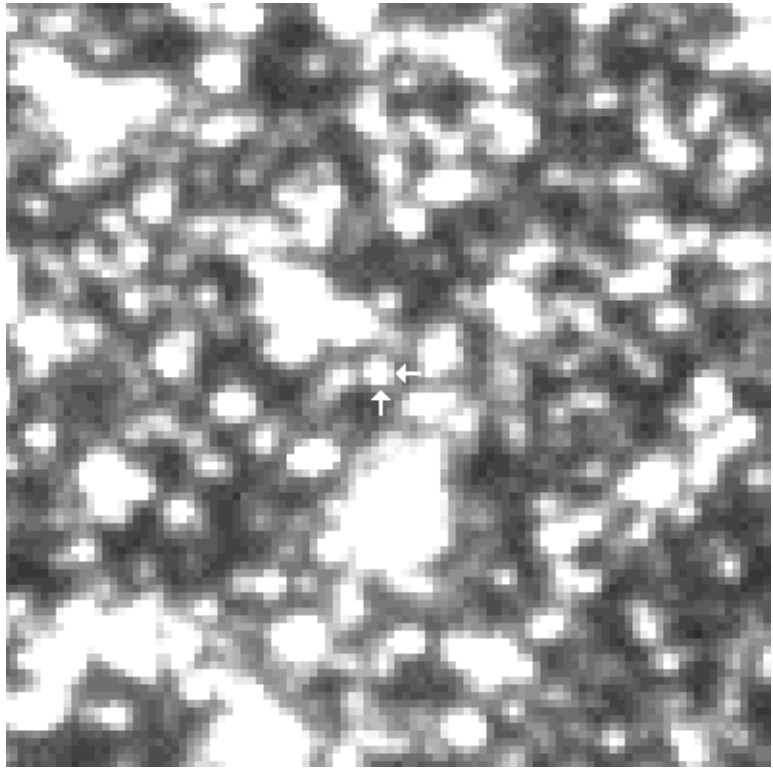
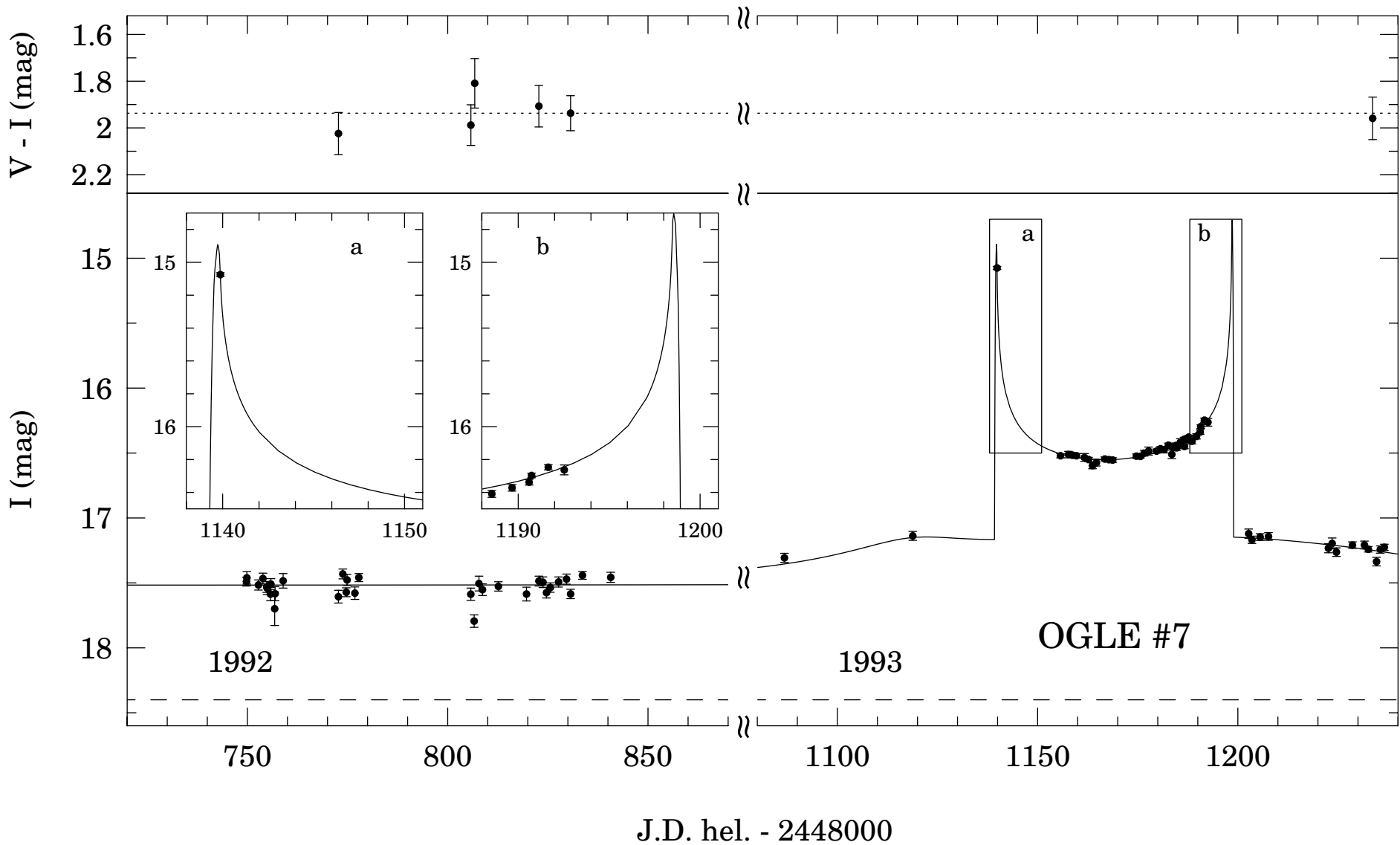
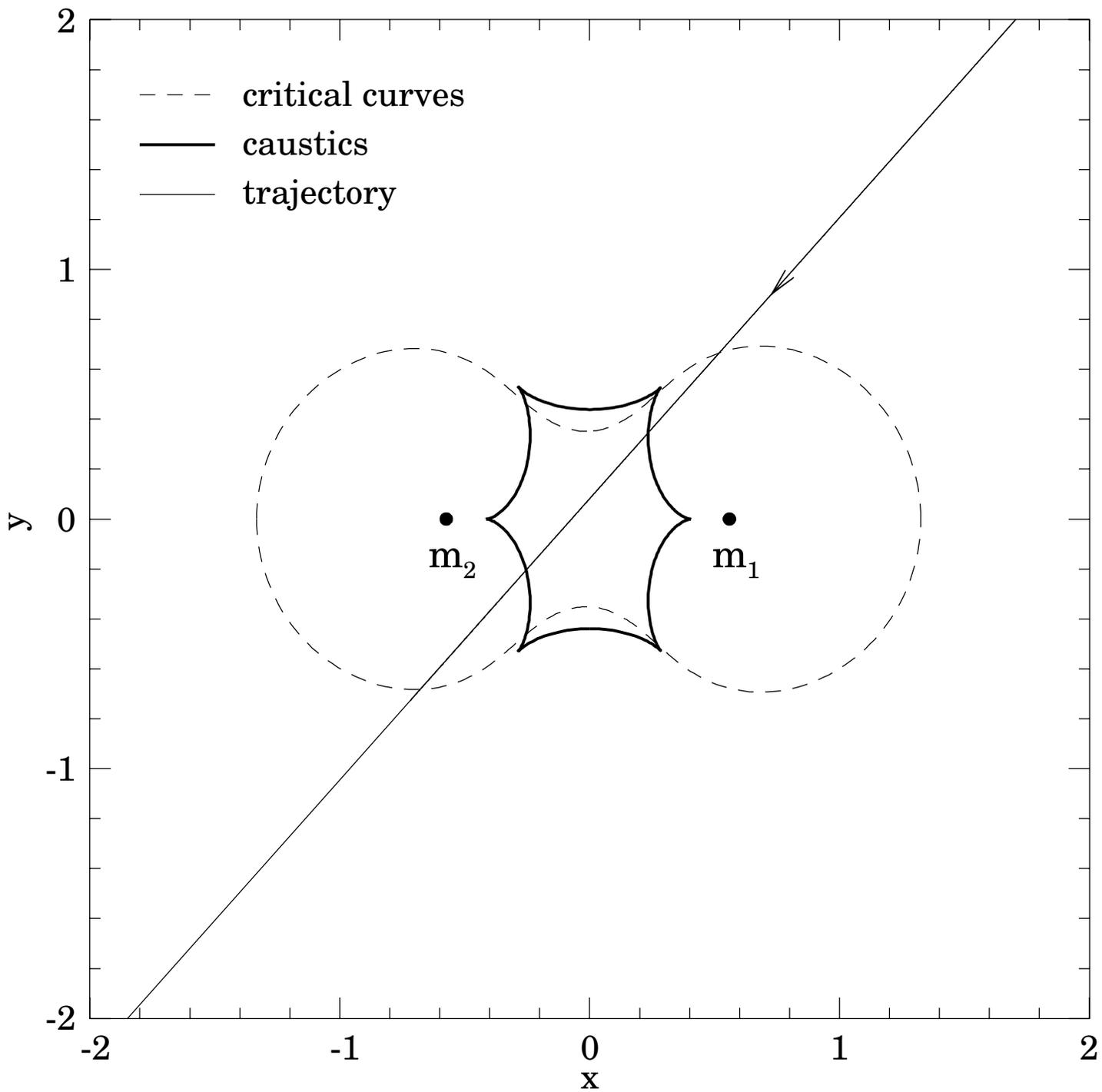
\end{document}